\documentclass[prb,preprint]{revtex4-1} 
\usepackage{amssymb}
\usepackage{color}
\usepackage{systeme,mathtools}

\usepackage{amsmath}  % needed for \tfrac, \bmatrix, etc.
\usepackage{amsfonts} % needed for bold Greek, Fraktur, and blackboard bold
\usepackage{graphicx} % needed for figures
\usepackage{systeme,mathtools}

\begin{document}

% Be sure to use the \title, \author, \affiliation, and \abstract macros
% to format your title page.  Don't use lower-level macros to  manually
% adjust the fonts and centering.

\title{A 3D printed wheel with constant mass and variable moment of inertia for lab and demonstration.}
% In a long title you can use \\ to force a line break at a certain location.

\author{Eric Hazlett,  }
\email{ehazlett@carleton.edu} % optional
%\altaffiliation[permanent address: ]{101 Main Street, 
 % Anytown, USA} % optional second address
% If there were a second author at the same address, we would put another 
\author{Andr\'es Aragoneses}
% \author{} statement here.  Don't combine multiple authors in a single
% \author statement.
\affiliation{Department of Physics and Astronomy, Carleton, Northfield, MN 55057}
% Please provide a full mailing address here.

% See the REVTeX documentation for more examples of author and affiliation lists.

\date{\today}

%\begin{abstract}
%We present a versatile experimental apparatus for exploring rotational motion through the interplay between the moment of inertia, torque and rotational kinetic energy.  The heart of this experiment uses a 3D printed wheel along with easily accessible stock components that allow for the adjustment of moment of inertial while keeping the total mass of the wheel constant. The wheel can act as a massive pulley of variable moment of inertia that allows students to measure the moment of inertia of the bare wheel by applying a constant torque to the system.  The wheel can also be used to explore rotational kinetic energy in the form of races down ramps.  The 3D printed aspect of this wheel allows anyone with access to a 3D printer to create, explore, and modify this wheel at a low cost allowing for more flexibility and accessibility for student and instructor exploration and modification.

%\end{abstract}
% AJP requires an abstract for all regular article submissions.
% Abstracts are optional for submissions to the "Notes and Discussions" section.

\maketitle % title page is now complete
{\large{
\begin{center}
\bf{}
\end{center}
}}
\section{Introduction.}
We present a versatile experimental apparatus for exploring rotational motion through the interplay between the moment of inertia, torque and rotational kinetic energy.  The heart of this experiment uses a 3D printed wheel along with easily accessible stock components that allow for the adjustment of moment of inertial while keeping the total mass of the wheel constant. The wheel can act as a massive pulley of variable moment of inertia that allows students to measure the moment of inertia of the bare wheel by applying a constant torque to the system.  The wheel can also be used to explore rotational kinetic energy in the form of races down ramps.  The 3D printed aspect of this wheel allows anyone with access to a 3D printer to create, explore, and modify this wheel at a low cost allowing for more flexibility and accessibility for student and instructor exploration and modification.

In the study of linear kinematics it is easy to investigate how systems evolve with variable mass.  With rotational motion the situation is more complicated due to the fact that the moment of inertia depends not only on the mass, but the distribution of the mass. This can be demonstrated in the form of massive pulley's in Atwood machines \cite{atwood} and in a rolling race of objects of varying mass and moment of inertia.  For lab explorations it is best to isolate a single variable that can be changed. For rotational motion is difficult to find a wheel with constant mass, but different moments of inertia.   There do exist examples that solve this problem \cite{spiderwheel,MoIInexpensive}, but they require workshop access and cannot act as pulleys. To address this issue we have designed a wheel whose moment of inertia can be easily created and manipulated while the mass of the system remains constant. Providing an accessible, flexible, and robust apparatus to explore the interplay between the moment of inertia and torque and rotational kinetic energy.  

The backbone of our apparatus is a 3D printed wheel with various location where steel dowel pins can be inserted.  3D printers are becoming more accessible in high schools, colleges and universities, and public libraries.  This eliminates the need to have access to a full machine shop for production.  The 3D printed design is open source and will hopefully evolve as improvement from students and instructors provide feedback.   Assuming access to posts and post clamps the experimental apparatus is affordable and accessible to classrooms everywhere.

\section{The apparatus}
\begin{figure}[tph]
\begin{centering}
  \includegraphics[width=6 in]{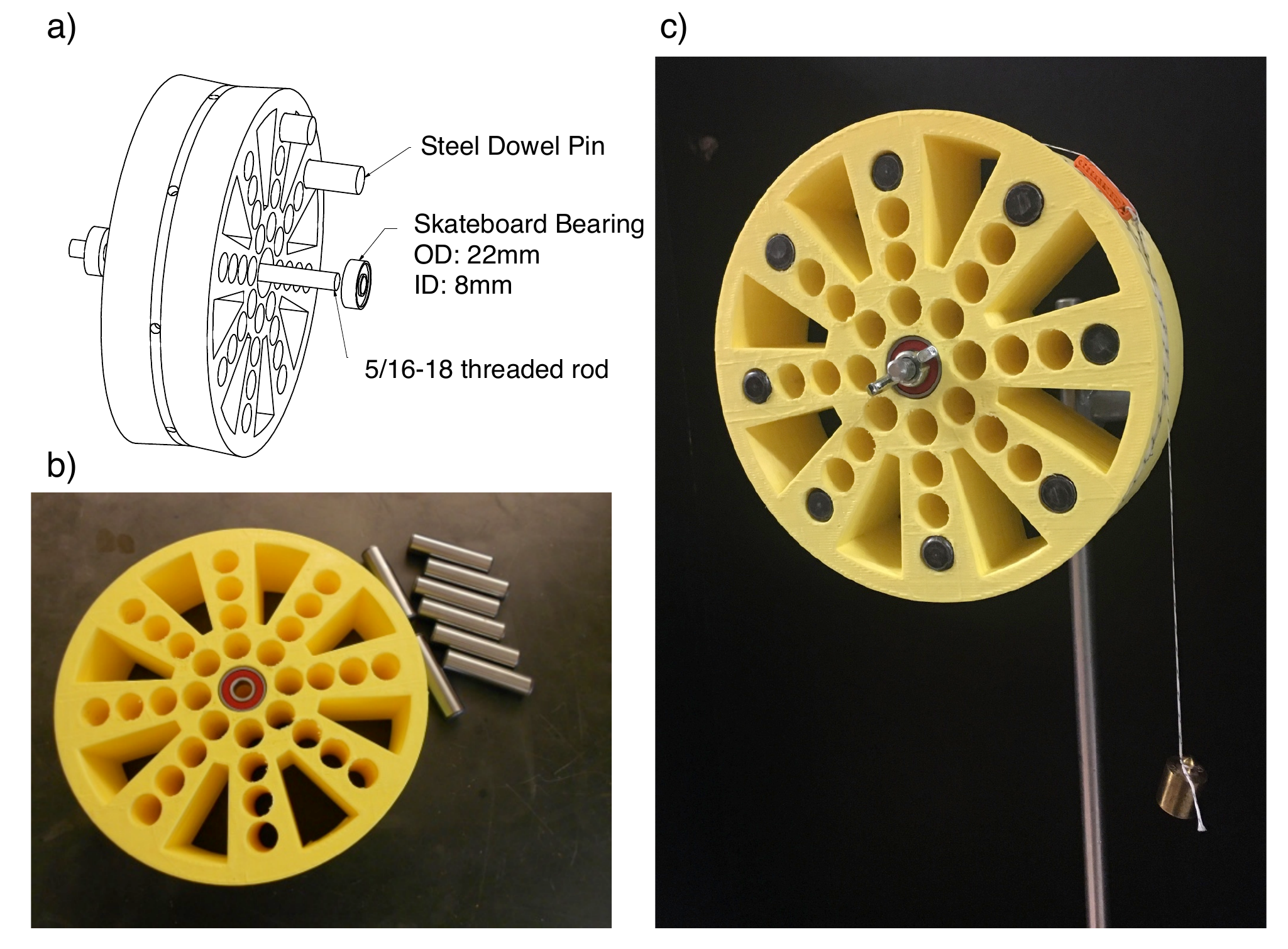}
     \caption{The wheel experiment for investigating moment of inertia.  a) Schematic drawing of the assembly of the wheel.  The dimensions are for parts that are consistent among different variations of the wheel. Not shown are the retaining nuts and washers as well as the stand for mounting the rod. b) Picture of the printed wheel with a thickness of 2 in, outside diameter of 6.5 in,  and the eight 0.5 in dowel pins. c) The full experimental setup showing the mounted wheel and the dowel pins inserted at the largest radial location. A 3D printed pin (orange in the figure) is used to secure the line in the mounting holes in the rim of the wheel. }
     \end{centering}
     \label{setup}
\end{figure}

The heart of the device is a 3D printed wheel with eight spokes where steel dowel pins can be inserted at various radii as shown in figure \ref{setup}.   For our investigation we used s $6.5$ inch outside radius and $0.5$ inch steel dowel pins that were 2 and 2.5 inches long. The length of the dowel pin does not need match the thickness of the wheel as long as a standard retaining nut is used on the threaded rod.  The wheel can be scaled to various sizes by modifying the design files.  At all scales there are 8 spokes evenly distributed angularly about the central axis. There are 4 radial positions in our configuration.  %The .ipt cad file used will automatically scale this for varying radius and dowel pin size. 
The radial edge of the wheel has a small taper that gives two contact points for rolling-cylinder investigations. The radial edge has some holes in it that allow for a string to be attached via a paper clip or 3D printed jig. At the center of each side there is a place for a 22 mm bearing that are standard skateboard bearings which comprise the hub of the unit.  Using a piece of 5/16 inch threaded rod for the axle the pulley is complete as shown in figure \ref{setup}.  All of the parts of this apparatus, along with detailed instructions on assembly, can be found on the Thingiverse repsoitory \cite{Thingiverse}.

The partlist and approximate unit cost  for one complete setup is:
\begin{itemize}
\item 1 --  3D printed wheel (ABS or PLA).  One wheel with a diameter of 6.5 in and a thickness of 2 in. $\sim$   \$15.
\item 8 -- Dowel pins:  1/2" diameter 2" length (for shown setup):  $\sim$ $8 \times \$1.8 =\$14.4$  
\item 2 -- Outside Diameter 22 mm, inside diamter 8 mm bearing (standard skateboard bearing):$\sim$  \$1.05 for two bearings.
\item 1 -- 5/16" or 8 mm threaded rod for mounting:  $\sim$ \$1.50
\item 2 -- mounting hardward $\sim$ \$1.

\end{itemize}
Using common post mounts and clamps to secure the threaded rod the per unit price of the apparatus becomes less that \$35.  Which could outfit an 8 team lab for less that \$200\footnote{Cost is a conservative estimated using PRO series PLA and ABS.  Buying in bulk also lowered cost}.  We also chose to buy the dowel pins in order to eliminate the need for a machine shop.  If standard grade and on hand materials are used then the cost can be pushed even lower to less that \$25 a setup or less depending on the setup.

\section{The experiment}

In our lab we first asked the students to calculate the rate of gravitational acceleration $g$ and moment of inertia $I_o$ of the wheel.  This is done by using the wheel as a pulley and varying the moment of inertia by redistribution of the masses.  For each configuration a small mass, $m_0$, is suspended from thread or fishing line that is wound around the pulley. By measuring the time and distance the mass falls, an accurate measurement of the acceleration $a$ can be found. When the mass is released and accelerates downward at $a$ given by Newton's second law,

\begin{equation}
%\sum \vec{F}_i = m_0a~~\Rightarrow ~
m_0g-T=m_0a.
\end{equation}

Where $g$ is the acceleration due to gravity, $a$ is the acceleration of the mass, $m_0$ is the suspended mass, and $T$ is the tension in the string. The acceleration can be found by measuring the time it takes for the released mass to fall a measured distance. The tension force also exerts a torque $\tau$ on the wheel such that 

\begin{equation}
T\cdot R=I_{tot}\alpha
\end{equation}

Where $R$ is the radius of the wheel, $I_{tot}$ is the total moment of inertia of the wheel and $\alpha$ is the angular acceleration of the wheel.  While the string remains taut the constraint that $a =\alpha R$ holds.  Using the parallel-axis theorem, $I_{tot}$ can be written in terms of the moment of inertia of the wheel, $I_{wheel}$, the moment of inertia of each dowel pin, $I_{p}=\frac{1}{2}m_1b^2$, and the radial location of each pin $r$;
\begin{equation}
I=I_{wheel}+\sum n_i \left[  \frac{1}{2}m_{p}b^2+m_{p}r_i^2 \right].
\end{equation}

Where $m_p$ is the mass of each pin, $b$ is the radius of the rods, and the sum is done for each radial position and $n_i$ is the number of rods at located at that radial position.  In our experiment we placed n pins at the same radius simplifying the above to be
\begin{equation}
I=I_{wheel}+ n\left[  \frac{1}{2}m_{p}b^2+m_{p}r^2 \right].
\end{equation}

After some algebra we get a relationship between the radial position of the rods and time in the form of

\begin{equation}
r^2 = \left[\frac{m_0gR^2}{2hnm_1}\right]t^2-\frac{m_0R^2}{nm_p}-\frac{I_{wheel}}{nm_p}-\frac{b^2}{2},
\label{result}
\end{equation}

where $h$ is the height the small mass was dropped over and $t$ the time it took to drop that distance. 

\begin{figure}[tph]
\begin{centering}
  \includegraphics[width=6 in]{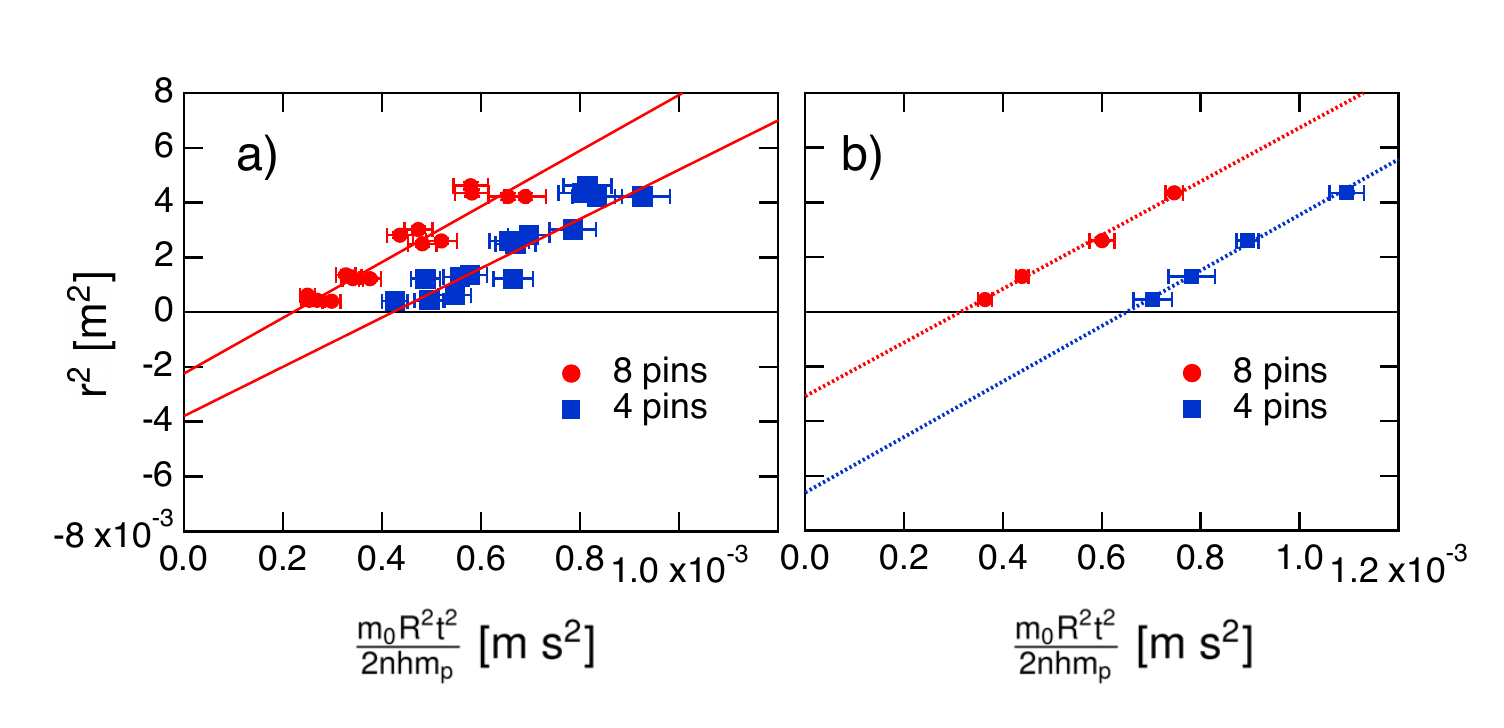}
     \caption{A linearized plot of our experimental data of a) five student groups and b) our trial runs. The student data used a drop mass of 10g and we used a drop mass of 20g.  The student's drop height was in the range of 1.5 m to 1.8 m and our drop height was 1.4 m.  The horizontal axis is scaled such that the slope yields a direct measurement of g. }
     \end{centering}
     \label{fit_figure}
\end{figure}

Equation \ref{result} can be linearized by plotting $r^2$ vs $t^2$ which yields the first goal of this lab.  From the linearization of Eq. \ref{result} the slope is proportional to measured quantities and $g$, which allows for a measurement of local acceleration due to gravity.  The intercept of the graph is proportional to the moment of inertia of the wheel.   The results are shown in Fig. \ref{fit_figure}. The average of the results obtained by the students are reasonable within experimental errors: $g=(10.16\pm0.34)~\frac{m}{s^2}$ for the 8 pin investigation and $g=(8.99\pm0.41)~\frac{m}{s^2}$ for the four pin investigation.  In our investigations we find that $g=(9.82\pm0.15)~\frac{m}{s^2}$ for the 8 pin investigation and $g=(10.15\pm0.19)~\frac{m}{s^2}$.   The moment of inertia is calculated from the intercept of the graph, the value the students obtained was $I_0=(1.01\pm0.08)\times10^{-3}~kg\cdot m^2$ and our value was $I_0=(1.13\pm0.04)\times10^{-3}~kg\cdot m^2$. The theoretical value for the moment of inertia of a solid (hollow) cylinder of the same dimensions and mass to the 3D printed one are $0.99\times10^{-3}~kg\cdot m^2$ and $1.98\times10^{-3}~kg\cdot m^2$, respectively, being the measured value between both, as expected.

This result provides the template for measuring the moment of inertia of arbitrary objects.  Now that the moment of inertia of the wheel is known an object of unknown moment of inertia can be attached to the wheel and the experiment performed again. 

The wheel also provides a demonstration of the rotational kinetic energy race \cite{RollingSphere}, where objects of equal masses, but different moment of inertia, roll down a ramp and the student is tasked to determine which object will finish first.  This shows the relationship between potential energy and translations and rotational kinetic energy.  Intuition on this subject is notoriously shaky \cite{StudentUnderstanding} , but this setup is a clear example for students to explore and understand the impact that moment of inertia plays.  

	While we outline one experiment in this report there this wheel could be used in a variety of different ways. With the variety of configurations of moment of inertia possible other challenges to the students can be raised such as "Can you have three wheels with different masses reach the bottom of the ramp at the same time?" Which could be answered analytically, computationally, and then experimentally determined with this setup. 
	
	  At the same time the 3D printed aspect of it allows the wheel to be easily modified by the students or instructor and have those modifications become reality without the need of a dedicated machinist or machine shop.  This experiment shows flexibility and utility of 3D printed apparatuses to explore concepts that are difficult for everyone to explore with traditional manufacturing processes. 	 
\bibliography{MoIBib}

\end{document}